# Universal Spectrum for DNA base CG Frequency Distribution in Takifugu rubripes (Puffer fish) Genome

## A. M. Selvam[1]

Deputy Director (Retired)
Indian Institute of Tropical Meteorology, Pune 411 008, India
Email: amselvam@gmail.com
Websites: http://www.geocities.ws/amselvam
http://amselvam.tripod.com/index.html

## Abstract

The frequency distribution of DNA bases A, C, G, T exhibit fractal fluctuations ubiquitous to dynamical systems in nature. The power spectra of fractal fluctuations exhibit inverse power law form signifying long-range correlations between local (small-scale) and global (large-scale) perturbations.

The author has developed a general systems theory based on classical statistical physics for fractal fluctuations which predicts that the probability distribution of eddy amplitudes and the variance (square of eddy amplitude) spectrum of fractal fluctuations follow the universal Boltzmann inverse power law expressed as a function of the golden mean. The model predicted distribution is very close to statistical normal distribution for fluctuations within two standard deviations from the mean and exhibits a fat long tail.

In this paper it is shown that DNA base CG frequency distribution in Takifugu rubripes (Puffer fish) Genome Release 4 exhibit universal inverse power law form consistent with model prediction. The observed long-range correlations in the DNA bases implies that the non-coding 'junk' or 'selfish' DNA which appear to be redundant, may also contribute to the efficient functioning of the protein coding DNA, a result supported by recent studies.

---

[1] Corresponding Author: *Present/permanent address*: B1 Aradhana, 42/2A Shivajinagar, Pune 411 005, India. Tel. (Res.): 091-022-25538194, *Email address*: amselvam@gmail.com

# Universal Spectrum for DNA base CG frequency Distribution in Takifugu rubripes (Puffer fish) Genome

## A. M. Selvam

Deputy Director (Retired)
Indian Institute of Tropical Meteorology, Pune 411 008, India
Email: amselvam@gmail.com
Websites: http://www.geocities.com/amselvam
http://amselvam.tripod.com/index.html

## Abstract

The frequency distribution of DNA bases A, C, G, T exhibit fractal fluctuations ubiquitous to dynamical systems in nature. The power spectra of fractal fluctuations exhibit inverse power law form signifying long-range correlations between local (small-scale) and global (large-scale) perturbations.

The author has developed a general systems theory based on classical statistical physics for fractal fluctuations which predicts that the probability distribution of eddy amplitudes and the variance (square of eddy amplitude) spectrum of fractal fluctuations follow the universal Boltzmann inverse power law expressed as a function of the golden mean. The model predicted distribution is very close to statistical normal distribution for fluctuations within two standard deviations from the mean and exhibits a fat long tail.

In this paper it is shown that DNA base CG frequency distribution in Takifugu rubripes (Puffer fish) Genome Release 4 exhibits universal inverse power law form consistent with model prediction. The observed long-range correlations in the DNA bases implies that the non-coding 'junk' or 'selfish' DNA which appear to be redundant, may also contribute to the efficient functioning of the protein coding DNA, a result supported by recent studies.

## 1. Introduction

The DNA bases A, C, G, T exhibit long-range spatial correlations manifested as inverse power law form for power spectra of spatial fluctuations [1]. Such non-local connections are intrinsic to the selfsimilar fractal space-time fluctuations exhibited by dynamical systems in nature, now identified as self-organized criticality, an intensive field of research in the new discipline of nonlinear dynamics and chaos. The physics of the observed self-organized criticality or the ubiquitous $1/f$ spectra found in many disciplines of study is not yet identified [2]. A recently developed general systems theory [3-6] predicts the observed non-local connections as intrinsic to quantumlike chaos governing the space-time evolution of dynamical systems in nature. Identification of long-range correlations in the frequency distribution of the bases A, C, G, T in the DNA molecule imply mutual communication and control between coding and non-coding DNA bases for cooperative organization of vital functions for the living system by the DNA molecule as a unified whole entity. In this paper it is shown that the power spectra of DNA bases CG concentration of Takifugu rubripes (Puffer fish) genome exhibits model predicted inverse power law form implying long-range correlations in the spatial distribution of the DNA bases CG concentration. Takifugu rubripes (Puffer fish) genome is about nine times smaller than the human genome with the same number of genes and therefore with less number of non-coding DNA and is therefore of special interest for identification of location of known and unknown genes in the human genome. The paper is organized as follows: Sec. 2 discusses a multidisciplinary approach for modeling biological complexity and summarises the general systems theory concepts and model predictions of universal properties characterizing the form and function of dynamical systems. In Sec. 3 it is shown that the probability distribution of eddy amplitudes and the variance (square of eddy amplitude) spectrum of fractal fluctuations follow the universal Boltzmann inverse power law expressed as a function of the golden mean. The model predicted distribution is very close to statistical normal distribution for fluctuations within two standard deviations from the mean and exhibits a fat long tail. In Sec. 4 it is shown that the above model predictions can be derived directly from basic concepts in classical statistical physics. Sec. 5 relates to current knowledge regarding DNA base A, C, G, T organization, its implication and the importance of Takifugu rubripes (Puffer fish) genome in relation to identification of genes and their location in the human genome. Sec. 6 gives details of the Takifugu rubripes (Puffer fish) DNA data sets and the analysis techniques used for the study. Sec. 7 contains the results and discussions of analysis of the data sets. Sec. 8 states the important conclusions of the present study, in particular, the role of the apparently redundant noncoding DNA which comprises more than 90% of the human genome.

## 1.1 Fractal fluctuations and statistical normal distribution

Statistical and mathematical tools are used for analysis of data sets and estimation of the probabilities of occurrence of events of different magnitudes in all branches of science and other areas of human interest. Historically, the statistical normal or the Gaussian distribution has been in use for nearly 400 years and gives a good estimate for probability of occurrence of the more frequent moderate sized events of magnitudes within two standard deviations from the mean. The Gaussian distribution is based on the concept of data independence, fixed mean and standard deviation with a majority of data events clustering around the mean. However, for real world infrequent hazardous extreme events of magnitudes greater than two standard deviations, the statistical normal distribution gives progressively increasing under-estimates of up to near zero probability. In the 1890s the power law or Pareto distributions with implicit long-range correlations were found to fit the fat tails exhibited by hazardous extreme events such as heavy rainfall, stock market crashes, traffic jams, the after-shocks following major earthquakes, etc. A historical review of statistical normal and the Pareto distributions are given by Andriani and McKelvey [7] and Selvam [8]. The spatial and/or temporal data sets in practice refer to real world or computed dynamical systems and are fractals with self-similar geometry and long-range correlations in space and/or time, i.e., the statistical properties such as the mean and variance are scale-dependent and do not possess fixed mean and variance and therefore the statistical normal distribution cannot be used to quantify/describe self-similar data sets. Though the observed power law distributions exhibit qualitative universal shape, the exact physical mechanism underlying such scale-free power laws is not yet identified for the formulation of universal quantitative equations for fractal fluctuations of all scales. In the following Sec. 3 the universal inverse power law for fractal fluctuations is shown to be a function of the golden mean based on general systems theory concepts for fractal fluctuations.

## 2. Multidisciplinary Approach for Modeling Biological Complexity

Computational biology involves extraction of the hidden patterns from huge quantities of experimental data, forming hypotheses as a result, and simulation based analyses, which tests hypotheses with experiments, providing predictions to be tested by further studies. Robust systems maintain their state and functions against external and internal perturbations, and robustness is an essential feature of biological systems. Structurally stable network configurations in biological systems increase insensitivity to parameter changes, noise and minor mutations [9]. Systems biology advocates a multidisciplinary approach for modeling biological complexity. Many features of biological complexity result from selforganization. Biological systems are, in general, global patterns produced by local interactions. One of the appealing aspects of the study of self-organized systems is that we do not need anything specific from biology to understand the existence of self-organization. Self-organization occurs for reasons that have to do with the organization of the interacting elements [10]. The first and most general criterion for systems thinking is the shift from the parts to the whole. Living systems are integrated wholes whose properties cannot be reduced to those of smaller parts [11]. Many disciplines may have helpful insights to offer or useful techniques to apply to a given problem, and to the extent that problem-focused research can bring together practitioners of different disciplines to work on shared problems, this can only be a good thing. A highly investigated but poorly understood phenomena, is the ubiquity of so-called $1/f$ spectra in many interesting phenomena, including biological systems [12].

## 2.1 General systems theory for fractal fluctuations in dynamical systems

The inverse power law form for power spectra of fractal fluctuations signifies an eddy continuum underlying the apparently irregular (unpredictable) fluctuation pattern. The fractal fluctuations may be visualized to result from the superimposition of a continuum of eddies (waves), the larger eddies enclosing the smaller eddies, i.e., the space-time integration of enclosed smaller eddies gives rise to formation of successively larger eddies. Such a simple concept of generation of large scale fluctuations from the integrated mean of inherent ubiquitous small-scale (turbulent) fluctuations gives the following equation [13] for the relationship between the eddy circulation speeds $W$ and $w*$ of large and turbulent eddies respectively and their corresponding radii $R$ and $r$.

$$W^2 = \frac{2}{\pi} \frac{r}{R} w_*^2 \qquad (1)$$

The above Eq. (1) represents the basic concepts underlying the general systems theory and is the governing equation for the growth of successively larger scale eddies resulting from the integrated mean of enclosed smaller scale eddies leading to the formation of an eddy continuum as explained in the following. The growth of the large eddy results from the cooperative existence of internal small scale eddies. Since the square of eddy circulation speeds $W^2$ and $w_*^2$ represent eddy energies (kinetic), the above equation also quantifies the ordered two-way energy flow between the larger and smaller scales in terms of the length scale ratio $z$ equal to $R/r$ and is independent of any other physical, chemical, electrical properties of the medium of propagation. Large eddy growth exhibits the complex dynamics of a fuzzy logic system which responds as a unified whole to a multitude of inputs. The signatures of internal smaller scale fluctuations are carried as fine scale structure of large eddy circulations and contribute to the long-term correlations or 'memory' exhibited by dynamical systems.

## 2.2 Fractals represent hierarchical communication networks

The evolution of dynamical systems is governed by ordered information communication between the small-scale internal networks and the overall large scale growth pattern. The hierarchical network architecture underlying fractal space-time fluctuations (Eq. 1) provides robust two-way information communication and control for integrity in the performance of vital functions specific to the dynamical system.

Complex networks from such different fields as biology, technology or sociology share similar organization principles. The possibility of a unique growth mechanism promises to uncover universal origins of collective behaviour. In particular, the emergence of self-similarity in complex networks raises the fundamental question of the growth process according to which these structures evolve [14].

## 2.3 Model Predictions

The general systems theory model predictions for the space-time fractal fluctuation pattern of dynamical systems [3-6, 8, 15], are given in the following

### 2.3.1 Quasicrystalline pattern for the network architecture

The power spectra of fractal fluctuations follow inverse power law form signifying an underlying eddy continuum structure. Visualization of large eddies as envelopes enclosing internal small scale eddies leads to the result that the successive eddy length/time scales of component eddies and their circulation speeds in the eddy continuum follow the Fibonacci mathematical number series such that the ratio of successive eddy radii $R_{n+1}/R_n$ and also circulation speeds $W_{n+1}/W_n$ is equal to the golden mean $\tau$ ($\approx 1.618$).

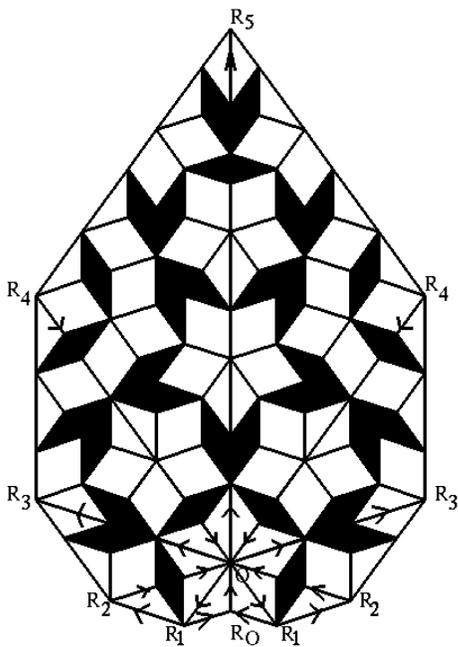

Fig. 1: Internal structure of large eddy circulations.

The apparently irregular fractal fluctuations can be resolved into a precise geometrical pattern with logarithmic spiral trajectory and the quasi periodic Penrose tiling pattern [16,17] for the internal structure (Fig. 1) on all scales to form a nested continuum of vortex roll circulations with ordered energy flow between the scales (Eq. 1). Logarithmic spiral formation with Fibonacci winding number and five-fold symmetry possess maximum packing efficiency for component parts and are manifested strikingly in plant Phyllotaxis [18,19].

Aperiodic or quasiperiodic order is found in different domains of science and technology. There is widespread presence of Fibonacci numbers and the golden mean in different physical contexts. Several conceptual links exist between quasiperiodic crystals and the hierarchical structure of biopolymers in connection with the charge transfer properties of both biological and synthetic DNA chains. DNA was originally viewed as a trivially periodic macromolecule, unable to store the amount of information required for the governance of cell function. The famous physicist, Schrodinger [20], was the first to suggest that genetic material should consist of a long sequence of a few repeating elements exhibiting a well-defined order without the recourse of periodic repetition, thereby introducing the notion of aperiodic crystal. Nevertheless, this notion remained dormant for almost four decades until the discovery of quasicrystalline solids. An essential characteristic of the quasicrystalline order is self-similarity which reveals the existence of certain motifs in the sample which contain the whole structure enfolded within them [21].

### 2.3.2 Fractal fluctuations signify quantum-like chaos

A hierarchy of logarithmic spiral circulations contributes to the formation of the observed self-similar fractal fluctuations in dynamical systems. The spiral flow structure $OR_OR_1R_2R_3R_4R_5$ can be visualized as an eddy continuum generated by successive length step growths $OR_O$, $OR_1$, $OR_2$, $OR_3$,....respectively equal to $R_1$, $R_2$, $R_3$,....which follow *Fibonacci* mathematical series such that $R_{n+1}=R_n+R_{n-1}$ and $R_{n+1}/R_n=\tau$ where $\tau$ is the *golden mean* equal to $(1+\sqrt{5})/2$ ($\approx 1.618$). Considering a normalized length step equal to 1 for the last stage of eddy growth, the successively decreasing radial length steps can be expressed as 1, $1/\tau$, $1/\tau^2$, $1/\tau^3$, ......The normalized eddy continuum comprises of fluctuation length scales 1, $1/\tau$, $1/\tau^2$, ........ The probability of occurrence is equal to $1/\tau$ and $1/\tau^2$ respectively for eddy length scale $1/\tau$ in any one or both rotational (clockwise and anti-clockwise) directions. Eddy fluctuation length of amplitude $1/\tau$ has a probability of occurrence equal to $1/\tau^2$ in both rotational directions, i.e., the square of eddy amplitude represents the probability of occurrence in the eddy continuum. Similar result is observed in the subatomic dynamics of quantum systems which are visualized to consist of the superimposition of eddy fluctuations in wave trains (eddy continuum).

Therefore, square of the eddy amplitude or the variance represents the probability. Such a result that the additive amplitudes of eddies, when squared, represent the probability densities is observed for the subatomic dynamics of quantum systems such as the electron or photon [22]. Townsend's [13] visualization of large eddy structure as quantified in Eq. (1) leads to the important result that the self-similar fractal fluctuations of atmospheric flows are manifestations of quantumlike chaos.

The square of the eddy amplitude $W^2$ represents the kinetic energy $E$ given as (from Eq. 1)

$$E = H\nu$$

where $\nu$ (proportional to $1/R$) is the frequency of the large eddy and $H$ is a constant equal to $\dfrac{2}{\pi}r_*w_*^2$ for growth of large eddies sustained by constant energy input proportional to $w_*^2$ from fixed primary small scale eddy fluctuations. Energy content of eddies is therefore similar to quantum systems which can possess only discrete quanta or packets of energy content $h\nu$ where $h$ is a universal constant of nature (Planck's constant) and $\nu$ is the frequency in cycles per second of the electromagnetic radiation.

### 2.3.3 Long-range spatiotemporal correlations (coherence)

The overall logarithmic spiral pattern enclosing the internal small-scale closed networks $OR_OR_1$, $OR_1R_2$, ... may be visualized as a continuous smooth rotation of the phase angle $\theta$ ($R_OOR_1$, $R_OOR_2$, ... etc.) with increase in period. The phase angle $\theta$ for each stage of growth is equal to $r/R$ and is proportional to the variance $W^2$ (Eq. 1), the variance representing the intensity of fluctuations. The phase angle gives a measure of coherence or correlation in space-time fluctuations of different length scales. The model predicted continuous smooth rotation of phase angle with increase in length scale associated with logarithmic spiral flow structure is analogous to Berry's phase [23,24] in the subatomic dynamics of quantum systems. Berry's phase has been identified in atmospheric flows [3-6].

### 2.3.4 Emergence of order and coherence in biology

Macroscale coherent structures emerge by space-time integration of microscopic domain fluctuations. Such a concept of the autonomous growth of hierarchical network continuum with ordered energy flow between the scales is analogous to Prigogine's [25] concept of the spontaneous emergence of order and organization out of apparent disorder and chaos through a process of self-organization.

The emergence of dynamical organization observed in physical and chemical systems should be of importance to biology [26]. Underlying the apparent complexity (of living matter), there are fundamental organisational principles based on physicochemical laws [27].

The general systems theory for coherent pattern formation summarized above may provide a model for biological complexity. General systems theory is a logical-mathematical field, the subject matter of which is the formulation and deduction of those principles which are valid for 'systems' in general, whatever the nature of their component elements or the relations or 'forces' between them [28-30].

### 2.3.5 Dominant length scales in the quasicrystalline spatial pattern

The eddy continuum underlying fractal fluctuations has embedded robust dominant wavebands $R_OOR_1$, $R_1R_2O$, $R_3R_2O$, $R_3R_4O$, …… with length (time) scales $T_D$ which are functions of the golden mean $\tau$ and the primary eddy energy perturbation length (time) scale $T_S$ such as the annual cycle of summer to winter solar heating in atmospheric flows. The dominant eddy length (time) scale for the $n^{th}$ dominant eddy is given as

$$T_D = T_S \left(2 + \tau\right)\tau^n \tag{2}$$

The successive dominant eddy length (time) scales for unit primary perturbation length (time) scale, i.e. $T_S = 1$, are given (Eq. 2) as 2.2, 3.6, 5.9, 9.5, 15.3, 24.8, 40.1, 64.9, …respectively for values of $n$ = -1, 0, 1, 2, 3, 4, 5, 6,….

Space-time integration of small-scale internal circulations results in robust broadband dominant length scales which are functions of the primary length scale $T_S$ alone and are independent of exact details (chemical, electrical, physical etc.) of the small-scale fluctuations. Such global scale spatial oscillations in the unified network are not affected appreciably by failure of localized micro-scale circulation networks [9].

Wavelengths (or periodicities) close to the model predicted values have been reported in weather and climate variability [3-6], prime number distribution [31], Riemann zeta zeros (non-trivial) distribution [32], Drosophila DNA base sequence [33], stock market economics [34], Human chromosome 1 DNA base sequence [35].

## 3. Universal spectrum for fractal fluctuations

## 3.1 Logarithmic spiral pattern underlying fractal fluctuations

The overall logarithmic spiral flow structure $OR_OR_1R_2R_3R_4R_5$ (Fig. 1) is given by the relation

$$W = \frac{w_*}{k} \ln z \tag{3}$$

In Eq. (3) the constant $k$ is the steady state fractional volume dilution of large eddy by inherent turbulent eddy fluctuations and $z$ is the length scale ratio $R/r$. The constant $k$ is equal to $1/\tau^2$ ($\cong 0.382$) and is identified as the universal constant for deterministic chaos in fluid flows. The steady state emergence of fractal structures is therefore equal to

$$\frac{1}{k} = \frac{WR}{w_* r} \cong 2.62 \qquad (4)$$

In Eq. (3), $W$ represents the standard deviation of eddy fluctuations, since $W$ is computed as the instantaneous r. m. s. (root mean square) eddy perturbation amplitude with reference to the earlier step of eddy growth. For two successive stages of eddy growth starting from primary perturbation $w_*$, the ratio of the standard deviations $W_{n+1}$ and $W_n$ is given from Eq. (3) as $(n+1)/n$. Denoting by $\sigma$ the standard deviation of eddy fluctuations at the reference level ($n=1$) the standard deviations of eddy fluctuations for successive stages of eddy growth are given as integer multiples of $\sigma$, i.e., $\sigma$, $2\sigma$, $3\sigma$, etc. and correspond respectively to

$$\text{statistical normalised standard deviation} \quad t = 0, 1, 2, 3, \ldots \qquad (5)$$

The conventional power spectrum plotted as the variance versus the frequency in log-log scale will now represent the eddy probability density on logarithmic scale versus the standard deviation of the eddy fluctuations on linear scale since the logarithm of the eddy wavelength represents the standard deviation, i.e., the r. m. s. value of eddy fluctuations (Eq. 3). The r. m. s. value of eddy fluctuations can be represented in terms of statistical normal distribution as follows. A normalized standard deviation $t=0$ corresponds to cumulative percentage probability density equal to 50 for the mean value of the distribution. Since the logarithm of the wavelength represents the r. m. s. value of eddy fluctuations the normalized standard deviation $t$ is defined for the eddy energy as

$$t = \frac{\log L}{\log T_{50}} - 1 \qquad (6)$$

In Eq. (6) $L$ is the time period (or wavelength) and $T_{50}$ is the period up to which the cumulative percentage contribution to total variance is equal to 50 and $t = 0$. $\log T_{50}$ also represents the mean value for the r. m. s. eddy fluctuations and is consistent with the concept of the mean level represented by r. m. s. eddy fluctuations. Spectra of time series of meteorological parameters when plotted as cumulative percentage contribution to total variance versus normalized deviation $t$ have been shown to follow closely the model predicted universal spectrum [36-39] which is identified as a signature of quantum-like chaos. Substituting for the golden mean $\tau$ equal 1.618 and the primary length scale $T_S$ equal to 1 in Eq. (2), $T_{50}$ is equal to 3.618 length scale units (Eq. 7).

$$T_{50} = T_S (2 + \tau)\tau^n = (2 + \tau)\tau^0 \approx 3.618 \qquad (7)$$

## 3.2 Universal Feigenbaum's constants and probability distribution function for fractal fluctuations

Selvam [5,40] has shown that Eq. (1) represents the universal algorithm for deterministic chaos in dynamical systems and is expressed in terms of the universal *Feigenbaum*'s *constants a* and *d* [41] as follows. The successive length step growths generating the eddy continuum $OR_OR_1R_2R_3R_4R_5$ (Fig. 1) analogous to the period doubling route to chaos (growth) is initiated and sustained by the turbulent (fine scale) eddy acceleration $w_*$, which then propagates by the inherent property of inertia of the medium of propagation. Therefore, the statistical parameters *mean*, *variance*, *skewness* and *kurtosis* of the perturbation field in the medium of propagation are given by $w_*, w_*^2, w_*^3$ and $w_*^4$ respectively. The associated dynamics of the perturbation field can be described by the following parameters. The perturbation speed $w_*$ (motion) per second (unit time) sustained by its inertia represents the mass, $w_*^2$ the acceleration or force, $w_*^3$ the angular momentum or potential energy, and $w_*^4$ the spin angular momentum, since an eddy motion has an inherent curvature to its trajectory.

It is shown that *Feigenbaum's* constant $a$ is equal to [5,40]

$$a = \frac{W_2 R_2}{W_1 R_1} \tag{8}$$

In Eq. (8) the subscripts 1 and 2 refer to two successive stages of eddy growth. *Feigenbaum's* constant $a$ as defined above represents the steady state emergence of fractional *Euclidean* structures. Considering dynamical eddy growth processes, *Feigenbaum's* constant $a$ also represents the steady state fractional outward mass dispersion rate and $a^2$ represents the energy flux into the environment generated by the persistent primary perturbation $W_1$. Considering both clockwise and counterclockwise rotations, the total energy flux into the environment is equal to $2a^2$. In statistical terminology, $2a^2$ represents the variance of fractal structures for both clockwise and counterclockwise rotation directions.

The probability of occurrence $P_{tot}$ of fractal domain $W_1 R_1$ in the total larger eddy domain $W_n R_n$ in any (irrespective of positive or negative) direction is equal to

$$P_{tot} = \frac{W_1 R_1}{W_n R_n} = \tau^{-2n}$$

Therefore the probability $P$ of occurrence of fractal domain $W_1 R_1$ in the total larger eddy domain $W_n R_n$ in any one direction (either positive or negative) is equal to

$$P = \left( \frac{W_1 R_1}{W_n R_n} \right)^2 = \tau^{-4n} \tag{9}$$

The *Feigenbaum's* constant $d$ is shown to be equal to [5,40]

$$d = \frac{W_2^4 R_2^3}{W_1^4 R_1^3} \tag{10}$$

Eq. (10) represents the fractional volume intermittency of occurrence of fractal structures for each length step growth. *Feigenbaum's* constant $d$ also represents the relative spin angular momentum of the growing large eddy structures as explained earlier.

Eq. (1) may now be written as

$$2 \frac{W^2 R^2}{w_*^2 (\mathrm{d}R)^2} = \pi \frac{W^4 R^3}{w_*^4 (\mathrm{d}R)^3} \tag{11}$$

In Eq. (11) $\mathrm{d}R$ equal to $r$ represents the incremental growth in radius for each length step growth, i.e., $r$ relates to the earlier stage of eddy growth.

The Feigenbaum's constant $d$ represented by $R/r$ is equal to

$$d = \frac{W^4 R^3}{w_*^4 r^3} \tag{12}$$

For two successive stages of eddy growth

$$d = \frac{W_2^4 R_2^3}{W_1^4 R_1^3} \tag{13}$$

From Eq. (1)

$$W_1^2 = \frac{2}{\pi} \frac{r}{R_1} w_*^2$$

$$W_2^2 = \frac{2}{\pi} \frac{r}{R_2} w_*^2 \tag{14}$$

Therefore

$$\frac{W_2^2}{W_1^2} = \frac{R_1}{R_2} \tag{15}$$

Substituting in Eq. (13)

$$d = \frac{W_2^4 R_2^3}{W_1^4 R_1^3} = \frac{W_2^2}{W_1^2} \frac{W_2^2 R_2^3}{W_1^2 R_1^3} = \frac{R_1}{R_2} \frac{W_2^2 R_2^3}{W_1^2 R_1^3} = \frac{W_2^2 R_2^2}{W_1^2 R_1^2} \tag{16}$$

The Feigenbaum's constant $d$ represents the scale ratio $R_2/R_1$ and the inverse of the Feigenbaum's constant $d$ equal to $R_1/R_2$ represents the probability $(Prob)_1$ of occurrence of length scale $R_1$ in the total fluctuation length domain $R_2$ for the first eddy growth step as given in the following

$$\left(Prob\right)_1 = \frac{R_1}{R_2} = \frac{1}{d} = \frac{W_1^2 R_1^2}{W_2^2 R_2^2} = \tau^{-4} \tag{17}$$

In general for the $n^{\text{th}}$ eddy growth step, the probability $(Prob)_n$ of occurrence of length scale $R_1$ in the total fluctuation length domain $R_n$ is given as

$$\left(Prob\right)_n = \frac{R_1}{R_n} = \frac{W_1^2 R_1^2}{W_n^2 R_n^2} = \tau^{-4n} \tag{18}$$

The above equation for probability $(Prob)_n$ also represents, for the $n^{\text{th}}$ eddy growth step, the following statistical and dynamical quantities of the growing large eddy with respect to the initial perturbation domain: (i) the statistical relative variance of fractal structures, (ii) probability of occurrence of fractal domain in either positive or negative direction, and (iii) the inverse of $(Prob)_n$ represents the organized fractal (fine scale) energy flux in the overall large scale eddy domain. Large scale energy flux therefore occurs not in bulk, but in organized internal fine scale circulation structures identified as fractals.

Substituting the *Feigenbaum's constants* $a$ and $d$ defined above (Eqs. 8 and 10), Eq. (11) can be written as

$$2a^2 = \pi d \tag{19}$$

In Eq. (19) $\pi d$, the relative volume intermittency of occurrence contributes to the total variance $2a^2$ of fractal structures.

In terms of eddy dynamics, the above equation states that during each length step growth, the energy flux into the environment equal to $2a^2$ contributes to generate relative spin angular momentum equal to $\pi d$ of the growing fractal structures. Each length step growth is therefore associated with a factor of $2a^2$ equal to $2\tau^4$ ($\cong 13.70820393$) increase in energy flux in the associated fractal domain. Ten such length step growths results in the formation of robust (self-sustaining) dominant bidirectional large eddy circulation $OR_OR_1R_2R_3R_4R_5$ (Fig. 1) associated with a factor of $20a^2$ equal to 137.08203 increase in eddy energy flux. This non-dimensional constant factor characterizing successive dominant eddy energy increments is analogous to the *fine structure* constant $\propto^{-1}$ observed in atomic spectra [42], where the spacing (energy) intervals between adjacent spectral lines is proportional to the non-dimensional *fine structure* constant equal to approximately 1/137. Further, the probability of $n^{\text{th}}$ length step eddy growth is given by $a^{-2n}$ ($\cong 6.8541^{-n}$) while the associated increase in eddy energy flux into the environment is equal to $a^{2n}$ ($\cong 6.8541^n$). Extreme events occur for large number of length step growths $n$ with small probability of occurrence and are associated with large energy release in the fractal domain. Each length step growth is associated with one-tenth of *fine structure constant* energy increment equal to $2a^2$ ($\propto^{-1}/10 \cong 13.7082$) for bidirectional eddy circulation, or equal to one-twentieth of *fine structure constant* energy increment equal to $a^2$ ($\propto^{-1}/20 \cong 6.8541$) in any one direction, i.e., positive or negative. The energy increase between two successive eddy length step growths may be expressed as a function of $(a^2)^2$, i.e., proportional to the square of the *fine structure constant* $\propto^{-1}$. In the spectra of many atoms, what appears with coarse observations to be a single spectral line proves, with finer observation, to be a group of two or more closely spaced lines. The spacing of these fine-structure lines relative to the coarse spacing in the spectrum is proportional to the square of *fine structure constant*, for which reason this combination is called the *fine-structure constant*. We now know that the significance of the *fine-structure constant* goes beyond atomic spectra [42].

It was shown at Eq. (4) (Sec. 3.1) above that the steady state emergence of fractal structures in fluid flows is equal to $1/k$ ($=\tau^2$) and therefore the *Feigenbaum's constant a* is equal to

$$a = \tau^2 = \frac{1}{k} = 2.62 \tag{20}$$

### 3.3 Universal Feigenbaum's constants and power spectra of fractal fluctuations

The power spectra of fluctuations in fluid flows can now be quantified in terms of universal *Feigenbaum's constant a* as follows.

The normalized variance and therefore the statistical probability distribution is represented in terms of Feigenbaum's constant $a$ in Eq. 9 as follows.

$$P = a^{-2t} \tag{21}$$

In Eq. (21) $P$ is the probability density corresponding to normalized standard deviation $t$. The graph of $P$ versus $t$ will represent the power spectrum. The slope $Sl$ of the power spectrum is equal to

$$Sl = \frac{\mathrm{d}P}{\mathrm{d}t} \approx -P \tag{22}$$

The power spectrum therefore follows inverse power law form, the slope decreasing with increase in $t$. Increase in $t$ corresponds to large eddies (low frequencies) and is consistent with observed decrease in slope at low frequencies in dynamical systems.

The probability distribution of fractal fluctuations (Eq. 18) is therefore the same as variance spectrum (Eq. 21) of fractal fluctuations.

The steady state emergence of fractal structures for each length step growth for any one direction of rotation (either clockwise or anticlockwise) is equal to

$$\frac{a}{2} = \frac{\tau^2}{2}$$

since the corresponding value for both direction is equal to $a$ (Eqs. 4 and 20 ).

The emerging fractal space-time structures have moment coefficient of kurtosis given by the fourth moment equal to

$$\left(\frac{\tau^2}{2}\right)^4 = \frac{\tau^8}{16} = 2.9356 \approx 3$$

The moment coefficient of skewness for the fractal space-time structures is equal to zero for the symmetric eddy circulations. Moment coefficient of kurtosis equal to 3 and moment coefficient of skewness equal to *zero* characterize the statistical normal distribution. The model predicted power law distribution for fractal fluctuations is close to the Gaussian distribution.

### 3.4 The power spectrum and probability distribution of fractal fluctuations are the same

The relationship between *Feigenbaum's constant a* and power spectra may also be derived as follows.

The steady state emergence of fractal structures is equal to the *Feigenbaum's constant a* (Eqs. 4 and 20). The relative variance of fractal structure which also represents the probability $P$ of occurrence of bidirectional fractal domain for each length step growth is then equal to $1/a^2$. The normalized variance $\frac{1}{a^{2n}}$ will now represent the statistical probability density for the $n^{\text{th}}$ step growth according to model predicted quantum-like chaos for fluid flows. Model predicted probability density values $P$ are computed as

$$P = \frac{1}{a^{2n}} = \tau^{-4n} \tag{23}$$

or

$$P = \tau^{-4t} \tag{24}$$

In Eq. (24) $t$ is the normalized standard deviation (Eq. 5) for values of $t \geq 1$ and $t \leq -1$. The model predicted $P$ values corresponding to normalized deviation $t$ values less than 2 are slightly less than the corresponding statistical normal distribution values while the $P$ values are noticeably larger for normalized deviation $t$ values greater than 2 (Table 1 and Fig. 2) and may explain the reported *fat tail* for probability distributions of various physical parameters [43].

Values of the normalized deviation $t$ in the range $-1 < t < 1$ refer to regions of primary eddy growth where the fractional volume dilution $k$ (Eq. 4) by eddy mixing process has to be taken into account for determining the probability distribution $P$ of fractal fluctuations (see Sec. 3.5 below).

## 3.5 Primary eddy growth region fractal space-time fluctuation probability distribution

Normalized deviation $t$ ranging from $-1$ to $+1$ corresponds to the primary eddy growth region. In this region the probability $P$ is shown to be equal to $P = \tau^{-4k}$ (see below) where $k$ is the fractional volume dilution by eddy mixing (Eq. 4).

The normalized deviation $t$ represents the length step growth number for growth stage more than one. The first stage of eddy growth is the primary eddy growth starting from unit length scale ($r = 1$) perturbation, the complete eddy forming at the tenth length scale growth, i.e., $R = 10r$ and scale ratio $z$ equals 10 [3]. The steady state fractional volume dilution $k$ of the growing primary eddy by internal smaller scale eddy mixing is given from Eq. (4) as

$$k = \frac{w_* r}{WR} \qquad (25)$$

The expression for $k$ in terms of the length scale ratio $z$ equal to $R/r$ is obtained from Eq. (1) as

$$k = \sqrt{\frac{\pi}{2z}} \qquad (26)$$

A fully formed primary large eddy length $R = 10r$ ($z=10$) represents the average or mean level zero and corresponds to a maximum of 50% probability of occurrence of either positive or negative fluctuation peak at normalized deviation $t$ value equal to zero by convention. For intermediate eddy growth stages, i.e., $z$ less than 10, the probability of occurrence of the primary eddy fluctuation does not follow conventional statistics, but is computed as follows taking into consideration the fractional volume dilution of the primary eddy by internal turbulent eddy fluctuations. Starting from unit length scale fluctuation, the large eddy formation is completed after 10 unit length step growths, i.e., a total of 11 length steps including the initial unit ($r = 1$) perturbation. At the second step ($z = R/r = 2$) of eddy growth the value of normalized deviation $t$ is equal to 1.1 - 0.2 (= 0.9) since the complete primary eddy length plus the first length step is equal to 1.1. The probability of occurrence of the primary eddy perturbation at this $t$ value however, is determined by the fractional volume dilution $k$ which quantifies the departure of the primary eddy from its undiluted average condition and therefore represents the normalized deviation $t$. Therefore the probability density $P$ of fractal fluctuations of the primary eddy is given using the computed value of $k$ (Eq. 26) as shown in the following equation.

$$P = \tau^{-4k} \qquad (27)$$

The probabilities of occurrence ($P$) of the primary eddy for a complete eddy cycle either in the positive or negative direction are given for progressive growth stages ($t$ values) in the following Table 1. The statistical normal probability density distribution corresponding to the normalized deviation $t$ values are also given in the Table 1. The model predicted probability density distribution $P$ along with the corresponding statistical normal distribution with probability values plotted on linear and logarithmic scales respectively on the left and right hand sides are shown in Fig. 2. The model predicted probability distribution $P$ for fractal space-time fluctuations is very close to the statistical normal distribution for normalized deviation $t$ values less than 2 as seen on the left hand side of Fig. 2. The model predicts progressively higher values of probability $P$ for values of $t$ greater than 2 as seen on a logarithmic plot on the right hand side of Fig. 2 and may explain the reported *fat tail* for probability distributions of various physical parameters [43].

Table 1: Primary eddy growth

| Growth step no $z$ | $\pm t$ | $k$ | Probability (%) | |
|---|---|---|---|---|
| | | | Model predicted | Statistical normal |
| 2 | .9000 | .8864 | 18.1555 | 18.4060 |
| 3 | .8000 | .7237 | 24.8304 | 21.1855 |
| 4 | .7000 | .6268 | 29.9254 | 24.1964 |
| 5 | .6000 | .5606 | 33.9904 | 27.4253 |
| 6 | .5000 | .5118 | 37.3412 | 30.8538 |
| 7 | .4000 | .4738 | 40.1720 | 34.4578 |
| 8 | .3000 | .4432 | 42.6093 | 38.2089 |
| 9 | .2000 | .4179 | 44.7397 | 42.0740 |
| 10 | .1000 | .3964 | 46.6250 | 46.0172 |
| 11 | 0 | .3780 | 48.3104 | 50.0000 |

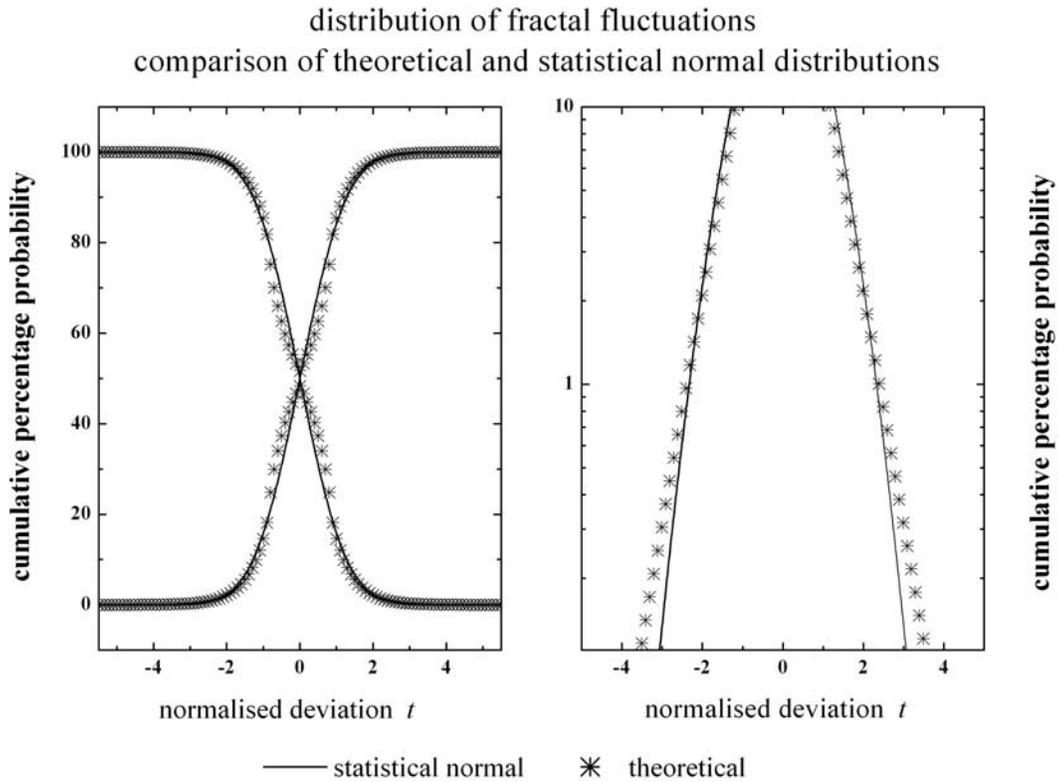

distribution of fractal fluctuations
comparison of theoretical and statistical normal distributions

Fig. 2: Probability distribution of fractal fluctuations. Comparison of theoretical with statistical normal distribution.

The probability distribution and the power (variance) spectrum of fractal fluctuations follow the same inverse power law $P = \tau^{-4t}$ where $P$ is the probability density and $t$ is the normalized deviation equal to $(av - x)/sd$ where $av$ and $sd$ are respectively the average and standard deviation of the fractal data series. The probability density $P$ also represents the normalized variance and corresponding normalized deviation $t$ equal to [log $(L)$/log $(T_{50})$ -1] where $L$ is the wavelength (period) and $T_{50}$ the wavelength (period) up to which the cumulative percentage contribution to total variance is equal to 50. The corresponding phase spectrum also follows the same inverse power law since eddy circulations are associated with phase angle equal to $r/R$ in Eq. (1) and represent the variance spectrum. Long-range space-time correlations are inherent to inverse power-law distributions. The model predicted probability density $P$ is very close to the statistical normal distribution for normalized deviation $t$ values less than 2, i.e. moderate amplitude fluctuations. For larger amplitude fluctuations, i.e., $t > 2$ the model predicted probability density $P$ is progressively larger than the corresponding statistical normal distribution. The applicability of statistical normal distribution for fractal fluctuations was discussed in Sec. 1.1 above.

The above model prediction, namely that the additive amplitudes of eddies when squared (variance) represent probability densities of eddy fluctuations (amplitudes) is exhibited by the sub-atomic dynamics of quantum systems such as the electron or photon. Therefore, fractal fluctuations exhibit quantum-like chaos.

## 4. General Systems Theory and Classical Statistical Physics

Nature has a hierarchical structure, with time, length and energy scales ranging from the submicroscopic to the supergalactic. Surprisingly it is possible and in many cases essential to discuss these levels independently—quarks are irrelevant for understanding protein folding and atoms are a distraction when studying ocean currents. Nevertheless, it is a central lesson of science, very successful in the past three hundred years, that there are no new fundamental laws, only new phenomena, as one goes up the hierarchy. Thus, arrows of explanations between different levels always point from smaller to larger scales, although the origin of higher level phenomena in the more fundamental lower level laws is often very far from transparent. Statistical Mechanics (SM) provides a framework for describing how well-defined higher level patterns or behavior may result from the non-directed activity of a multitude of interacting lower level individual entities. The subject was developed for, and has had its greatest success so far in, relating mesoscopic and macroscopic thermal phenomena to the microscopic world of atoms and molecules. Statistical mechanics explains how macroscopic phenomena originate in the cooperative behavior of these microscopic particles [44].

The general systems theory visualizes the self-similar fractal fluctuations to result from a hierarchy of eddies, the larger scale being the space-time average of enclosed smaller scale eddies (Eq. 1) assuming constant values for the characteristic length scale $r$ and circulation speed $w_*$ throughout the large eddy space-time domain. The collective behavior of the ordered hierarchical eddy ensembles is manifested as the apparently irregular fractal fluctuations with long-range space-time correlations generic to dynamical systems. The concept that aggregate averaged eddy ensemble properties represent the eddy continuum belongs to 19[th] century classical statistical physics where the study of the properties of a system is reduced to a determination of average values of the physical quantities that characterize the state of the system as a whole [45] such as gases, e.g., the gaseous envelope of the earth, the atmosphere.

In classical statistical physics *kinetic theory of ideal gases* is a study of systems consisting of a great number of molecules, which are considered as bodies having a small size and mass [45-52] are employed to estimate average values of quantities characterizing aggregate molecular motion such as mean velocity, mean energy etc., which determine the macro-scale characteristics of gases. The mean properties of ideal gases are calculated with the following assumptions. (1) The intra-molecular forces are completely absent instead of being small. (2) The dimensions of molecules are ignored, and considered as material points. (3) The above assumptions imply the molecules are completely free, move rectilinearly and uniformly as if no forces act on them. (4) The ceaseless chaotic movements of individual molecules obey Newton's laws of motion.

The Austrian physicist Ludwig Boltzmann suggested that knowing the probabilities for the particles to be in any of their various possible configurations would enable to work out the overall properties of the system. Going one step further, he also made a bold and insightful guess about these probabilities - that any of the many conceivable configurations for the particles would be equally probable. Boltzmann's idea works, and has enabled physicists to make mathematical models of thousands of real materials, from simple crystals to superconductors. It reflects the fact that many quantities in nature - such as the velocities of molecules in a gas - follow "normal" statistics. That is, they are closely grouped around the average value, with a "bell curve" distribution. Boltzmann's guess about equal probabilities only works for systems that have settled down to equilibrium, enjoying, for example, the same temperature throughout. The theory fails in any system where destabilizing external sources of energy are at work, such as the haphazard motion of turbulent fluids or the fluctuating energies of cosmic rays. These systems don't follow normal statistics, but another pattern instead [53].

Cohen [54] discusses Boltzmann's equation as follows. In 1872 when Boltzmann derived in his paper: *Further studies on thermal equilibrium between gas molecules* [55], what we now call the Boltzmann equation, he used, following Clausius and Maxwell, the assumption of 'molecular chaos', and he does not seem to have realized the statistical, i.e., probabilistic nature of this assumption, i.e., of the assumption of the independence of the velocities of two molecules which are going to collide. He used both a dynamical and a statistical method. However, Einstein strongly disagreed with Boltzmann's statistical method, arguing that a statistical description of a system should be based on the dynamics of the system. This opened the way, especially for complex systems, for other than Boltzmann statistics. It seems that perhaps a combination of dynamics and statistics is necessary to describe systems with complicated dynamics [54]. Sornette [56] discusses the ubiquity of observed power law distributions in complex systems as follows. The extension of Boltzmann's distribution to out-of-equilibrium systems is the subject of intense scrutiny. In the quest to characterize complex systems, two distributions have played a leading role: the normal (or Gaussian) distribution and the power law distribution. Power laws obey the symmetry of scale invariance. Power law distributions and more generally regularly varying distributions remain robust functional forms under a large number of operations, such as linear combinations, products, minima, maxima, order statistics, powers, which may also explain their ubiquity and attractiveness. Research on the origins of power law relations, and efforts to observe and validate them in the real world, is extremely active in many fields of modern science, including physics, geophysics, biology, medical sciences, computer science, linguistics, sociology, economics and more. Power law distributions incarnate the notion that extreme events are not exceptional. Instead, extreme events should be considered as rather frequent and part of the same organization as the other events [56].

In the following it is shown that the general systems theory concepts are equivalent to Boltzmann's postulates and the *Boltzmann distribution* with the inverse power law expressed as a function of the golden mean is the universal probability distribution function for the observed fractal fluctuations which corresponds closely to statistical normal distribution for moderate amplitude fluctuations and exhibit a fat long tail for hazardous extreme events in dynamical systems.

For any system large or small in thermal equilibrium at temperature $T$, the probability $P$ of being in a particular state at energy $E$ is proportional to $e^{-\frac{E}{K_B T}}$ where $K_B$ is the *Boltzmann's constant*. This is called the *Boltzmann distribution* for molecular energies and may be written as

$$P \propto e^{-\frac{E}{K_B T}} \tag{28}$$

The basic assumption that the space-time average of a uniform distribution of primary small scale eddies results in the formation of large eddies is analogous to Boltzmann's concept of equal probabilities for the microscopic components of the system [53]. The physical concepts of the general systems theory (Sec. 2) enable to derive [57] *Boltzmann distribution* as shown in the following.

The r.m.s circulation speed $W$ of the large eddy follows a logarithmic relationship with respect to the length scale ratio $z$ equal to $R/r$ (Eq. 3 ) as given below

$$W = \frac{w_*}{k} \log z$$

In the above equation the variable $k$ represents for each step of eddy growth, the fractional volume dilution of large eddy by turbulent eddy fluctuations carried on the large eddy envelope [3] and is given as (Eq. 25)

$$k = \frac{w_* r}{WR}$$

Substituting for $k$ in Eq. (3) we have

$$W = w_* \frac{WR}{w_* r} \log z = \frac{WR}{r} \log z$$

*and* $\tag{29}$

$$\frac{r}{R} = \log z$$

The ratio $r/R$ represents the fractional probability $P$ of occurrence of small-scale fluctuations ($r$) in the large eddy ($R$) environment. Since the scale ratio $z$ is equal to $R/r$ (Eq. 29) may be written in terms of the probability $P$ as follows.

$$\frac{r}{R} = \log z = \log\left(\frac{R}{r}\right) = \log\left(\frac{1}{(r/R)}\right)$$

$$P = \log\left(\frac{1}{P}\right) = -\log P \tag{30}$$

The maximum entropy principle concept of classical statistical physics is applied to determine the fidelity of the inverse power law probability distribution $P$ (Eq. 24) for exact quantification of the observed space-time fractal fluctuations of dynamical systems ranging from the microscopic dynamics of quantum systems to macro-scale real world systems. Kaniadakis [58] states that the correctness of an analytic expression for a given power-law tailed distribution, used to describe a statistical system, is strongly related to the validity of the generating mechanism. In this sense the maximum entropy principle, the cornerstone of statistical physics, is a valid and powerful tool to explore new roots in searching for generalized statistical theories [58]. The concept of entropy is fundamental in the foundation of statistical physics. It first appeared in thermodynamics through the second law of thermodynamics. In statistical mechanics, we are interested in the disorder in the distribution of the system over the permissible microstates. The measure of disorder first provided by Boltzmann principle (known as Boltzmann entropy) is given by $S = K_B \ln M$, where $K_B$ is the thermodynamic unit of measurement of entropy and is known as Boltzmann constant equal to $1.33 \times 10^{-16}$ erg/$^{\circ}$C. The variable $M$, called thermodynamic probability or statistical weight, is the total number of microscopic complexions compatible with the macroscopic state of the system and corresponds to the "degree of disorder" or 'missing information' [59]. For a probability distribution among a discrete set of states the generalized entropy for a system out of equilibrium is given as [59-62]

$$S = -\sum_{j=1}^{n} P_j \ln P_j \tag{31}$$

In Eq. (31) $P_j$ is the probability for the $j^{th}$ stage of eddy growth in the present study and the entropy $S$ represents the 'missing information' regarding the probabilities. Maximum entropy $S$ signifies minimum preferred states associated with scale-free probabilities.

The validity of the probability distribution $P$ (Eq. 24) is now checked by applying the concept of maximum entropy principle [58]. Substituting for $\log P_j$ (Eq. 30) and for the probability $P_j$ in terms of the golden mean $\tau$ derived earlier (Eq. 24) the entropy $S$ is expressed as

$$S = -\sum_{j=1}^{n} P_j \log P_j = \sum_{j=1}^{n} P_j^2 = \sum_{j=1}^{n} \left( \tau^{-4n} \right)^2$$
$$S = \sum_{j=1}^{n} \tau^{-8n} \approx 1 \text{ for large } n \tag{32}$$

In Eq. (32) $S$ is equal to the square of the cumulative probability density distribution and it increases with increase in $n$, i.e., the progressive growth of the eddy continuum and approaches 1 for large $n$. According to the second law of thermodynamics, increase in entropy signifies approach of dynamic equilibrium conditions with scale-free characteristic of fractal fluctuations and hence the probability distribution $P$ (Eq. 24) is the correct analytic expression quantifying the eddy growth processes visualized in the general systems theory.

In the following it is shown that the eddy continuum energy distribution $P$ (Eq. 24) is the same as the *Boltzmann distribution* for molecular energies. From Eq. (29)

$$z = \frac{R}{r} = e^{\frac{r}{R}}$$

$$or \tag{33}$$

$$\frac{r}{R} = e^{-\frac{r}{R}}$$

The ratio $r/R$ represents the fractional probability $P$ (Eq. 24) of occurrence of small-scale fluctuations ($r$) in the large eddy ($R$) environment. Considering two large eddies of radii $R_1$ and $R_2$ ($R_2$ greater than $R_1$) and corresponding r.m.s circulation speeds $W_1$ and $W_2$ which grow from the same primary small-scale eddy of radius $r$ and r.m.s circulation speed $w_*$ we have from Eq. (1)

$$\frac{R_1}{R_2} = \frac{W_2^2}{W_1^2}$$

From Eq. (33)

$$\frac{R_1}{R_2} = e^{-\frac{R_1}{R_2}} = e^{-\frac{W_2^2}{W_1^2}} \tag{34}$$

The square of r.m.s circulation speed $W^2$ represents eddy kinetic energy. Following classical physical concepts [46] the primary (small-scale) eddy energy may be written in terms of the eddy environment temperature $T$ and the *Boltzmann's constant $K_B$* as

$$W_1^2 \propto K_B T \tag{35}$$

Representing the larger scale eddy energy as $E$

$$W_2^2 \propto E \tag{36}$$

The length scale ratio $R_1/R_2$ therefore represents fractional probability $P$ (Eq. 24) of occurrence of large eddy energy $E$ in the environment of the primary small-scale eddy energy $K_B T$ (Eq. 35). The expression for $P$ is obtained from Eq. (34) as

$$P \propto e^{-\frac{E}{K_B T}} \tag{37}$$

The above is the same as the *Boltzmann's equation* (Eq. 28).

The derivation of *Boltzmann's equation* from general systems theory concepts visualises the eddy energy distribution as follows: (1) The primary small-scale eddy represents the molecules whose eddy kinetic energy is equal to $K_B T$ as in classical physics. (2) The energy pumping from the primary small-scale eddy generates growth of progressive larger eddies [3]. The r.m.s circulation speeds $W$ of larger eddies are smaller than that of the primary small-scale eddy (Eq. 1). (3) The space-time *fractal* fluctuations of molecules (atoms) in an ideal gas may be visualized to result from an eddy continuum with the eddy energy $E$ per unit volume relative to primary molecular kinetic energy ($K_B T$) decreasing progressively with increase in eddy size.

The eddy energy probability distribution ($P$) of fractal space-time fluctuations also represents the *Boltzmann distribution* for each stage of hierarchical eddy growth and is given by Eq. (24) derived earlier, namely

$$P = \tau^{-4t}$$

The general systems theory concepts are applicable to all space-time scales ranging from microscopic scale quantum systems to macroscale real world systems such as atmospheric flows.

## 5. DNA sequence and functions

The double-stranded DNA molecule is held together by chemical components called bases; Adenine (A) bonds with thymine (T); cytosine (C) bonds with guanine (G) These letters form the "code of life"; there are close to 3 billion base pairs in mammals such as humans and rodents. Written in the DNA of these animals are 25,000-30,000 genes which cells use as templates to start the production of proteins; these sophisticated molecules build and maintain the body. According to the traditional viewpoint, the really crucial things were genes, which code for proteins - the "building blocks of life". A few other sections that regulate gene function were also considered useful. The rest was thought to be excess baggage - or "junk" DNA. But new findings suggest this interpretation may be wrong. Comparison of genome sequences of man, mouse and rat and also chicken, dog and fish sequences show that several great stretches of DNA were identical across the species which shared an ancestor about 400 million years ago. These "ultra-conserved" regions do not appear to code for protein, but obviously are of great importance for survival of the animal. Nearly a quarter of the sequences overlap with genes and may help in protein production. The conserved elements that do not actually overlap with genes tend to cluster next to genes that play a role in embryonic development [63,64].

A number of the elements in the noncoding portion of the genome may be central to the evolution and development of multicellular organisms. Understanding the functions of these non-protein-coding sequences, not just the functions of the proteins themselves, will be vital to understanding the genetics, biology and evolution of complex organisms [65].

Brenner et al. [66] proposed as a model for genomic studies the tiger pufferfish (Taki) Fugu rubripes, a marine pufferfish with a genome nine times more compact than that of human. Fugu rubripes is separated from Homo sapiens by about 450 million years of evolution. Many comparisons have been made between F. rubripes and human DNA that demonstrate the potential of comparative genomics using the pufferfish genome [67]. Fugu's genome is compact, with less than 20% of repetitive sequences and fully one third occupied by gene loci. Aparicio et al. [68] report the sequencing and initial analysis of the Fugu genome. The usefulness of vertebrate comparative genomics was demonstrated by the identification of about thousand new genes in the human genome through comparison with Fugu. Functional comparisons between different fish genomes and the genomes of higher vertebrates will shed new light on vertebrate evolution and lead to the identification of the genes that distinguish fish and humans.

In the present study it is shown that long-range spatial correlations are exhibited by *Fugu rubripes* DNA base C+G concentration per 10 bp (base pair) along different lengths of DNA containing both coding and non-coding sequences indicating global sequence control over coding functions.

The study of C+G variability in genomic DNA is of special interest because of the documented association of gene dense regions in GC rich DNA sequences. GC-rich regions include many genes with short introns while GC-poor regions are essentially deserts of genes. This suggests that the distribution of GC content in mammals could have some functional relevance, raising the issue of its origin and evolution [69,70].

## 6. Data and Analysis

### 6.1 Data

The draft sequence of Takifugu rubripes (Puffer fish) genome assembly release 4 was obtained from "The Fugu Informatics Network" (ftp://fugu.biology.qmul.ac.uk/pub/fugu/scaffolds_4.zip) at School of Biological & Chemical Sciences, Queen Mary, University of London. The fourth assembly of the Fugu genome consists of 7,000 scaffolds. The individual contigs sizes range from 2-1100Kbp, nearly half the genome in just 100 scaffolds, 80% of the genome in 300 scaffolds.

Non-overlapping DNA sequence lengths without breaks (N values) were chosen and then grouped in two categories A and B; category A consists of DNA lengths greater than 3Kbp (kilo base pairs) but less than 30Kbp and category B consists of DNA sequence lengths greater than 30Kbp. The data series were subjected to continuous periodogram power spectral analyses. For convenience in averaging the results, the Category A data lengths were grouped into 7 groups and category B data into 18 groups such that the total number of bases in each group is equal to about 12Mbp

### 6.2 Power spectral analyses: variance and phase spectra

The number of times base C and also base G, i.e., (C+G), occur in successive blocks of 10 bases were determined in the DNA length sections giving a C+G frequency distribution series of 300 to 3000 values in category A and more than 3000 for category B. The power spectra of frequency distribution of C+G bases (per 10bp) in the data sets were computed accurately by an elementary, but very powerful method of analysis developed by Jenkinson [71] which provides a quasi-continuous form of the classical periodogram allowing systematic allocation of the total variance and degrees of freedom of the data series to logarithmically spaced elements of the frequency range (0.5, 0). The cumulative percentage contribution to total variance was computed starting from the high frequency side of the spectrum. The power spectra were plotted as cumulative percentage contribution to total variance versus the normalized standard deviation $t$ equal to $\left(\log L / \log T_{50}\right) - 1$ where $L$ is the period in years and $T_{50}$ is the period up to which the cumulative percentage contribution to total variance is equal to 50 (Eq. 6). The corresponding phase spectra were computed as the cumulative percentage contribution to total rotation (Section 2.3.3). The statistical chi-square test [72] was applied to determine the 'goodness of fit' of variance and phase spectra with statistical normal distribution. The variance and phase spectra were considered to be the same as model predicted spectrum for regions ($t$-values) where the model predicted spectrum lies within two standard deviations from mean variance and phase spectrum.

## 7 Results and discussion

### 7.1 Data sets and power spectral analyses

The average, standard deviation, maximum, minimum and median DNA lengths (bp) for each group in the two data categories A and B are shown in Fig. 3. It is seen that the mean is close to the median and almost constant for the different data groups, particularly for category B (DNA length > 30kbp).

Fig. 4 gives for each group in the two categories A and B, (i) the average and standard deviation of CG density per 10 bp (ii) the average and standard deviation of $T_{50}$ in length unit 10bp, (iii) the percentage number of variance spectra (V) and phase spectra (P) following normal distribution and the number of data sets.

The average CG density per 10bp is about 47% and 45% respectively for categories A and B data groups. Elgar et al. [73] report a G+C content of 47.67% and state that it is significantly higher than mammalian figures of 40.3% [74] and may reflect the higher proportion of coding sequence in the Fugu genome [66]. The higher CG density in Puffer fish genome may be attributed to less number of noncoding DNA as compared to mammalian genome. The noncoding DNA regions are associated with less CG concentration as compared to the coding regions.

The length scales upto which the cumulative percentage contribution to total variance is equal to 50 has been denoted as $T_{50}$ (Eq. 6) and the computed values obtained from spectral analyses are equal to about 6.4 and 6 in length unit 10bp, i.e., 64 and 60 bp, respectively for categories A and B data groups. These computed values are about twice the model predicted value equal to 36 bp for unit primary DNA length scale $OR_O = T_S = 10$ bp (Fig. 1). The observed higher value may be explained as follows. The first level of organized structure in the DNA molecule is the nucleosome formed by two turns of about 146 bp around a histone protein octamer [75,76]. Each turn of the nucleosome contains 73 bp and may represent a model predicted (Eq. 4) $T_{50} \approx 60$ bp for fundamental length scale $T_S = 10\tau$ bp$=OR_1$, (Fig. 1). A single turn of the nucleosome structure, the primary level of stable organized unit of the DNA molecule contributes upto 50 percent of the total variance signifying its role as internal structure which makes up the larger scale architecture of the DNA molecule.

Spectral analyses - DNA base CG frequency distribution
Takifugu rubripes (Puffer fish) Release 4

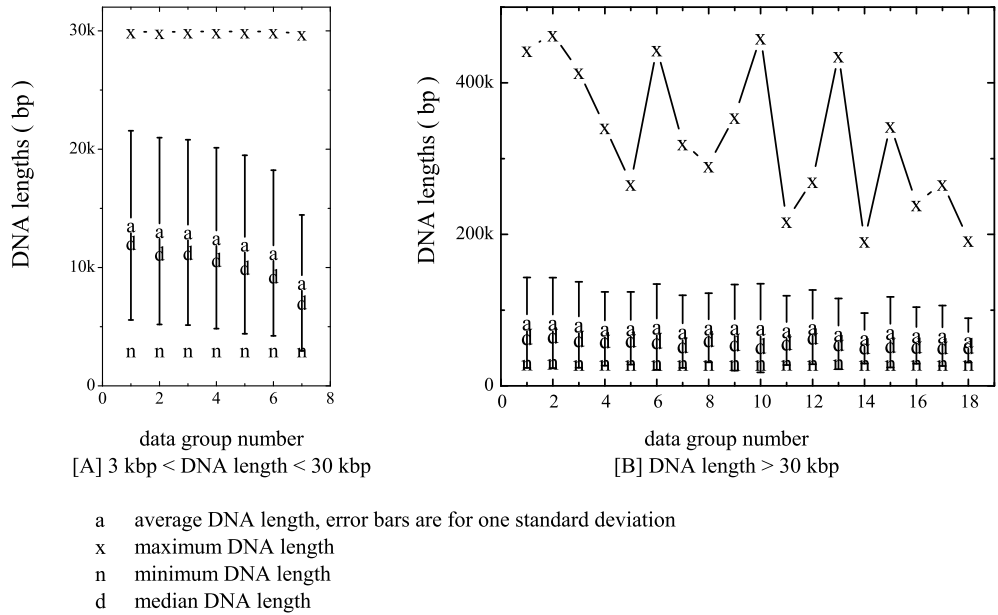

[A] 3 kbp < DNA length < 30 kbp          [B] DNA length > 30 kbp

a    average DNA length, error bars are for one standard deviation
x    maximum DNA length
n    minimum DNA length
d    median DNA length

Fig. 3. Details of data sets used for analyses

Spectral analyses DNA base CG frequency distribution
Takifugu rubripes (Puffer fish) Release 4 - Details of data sets and averaged results

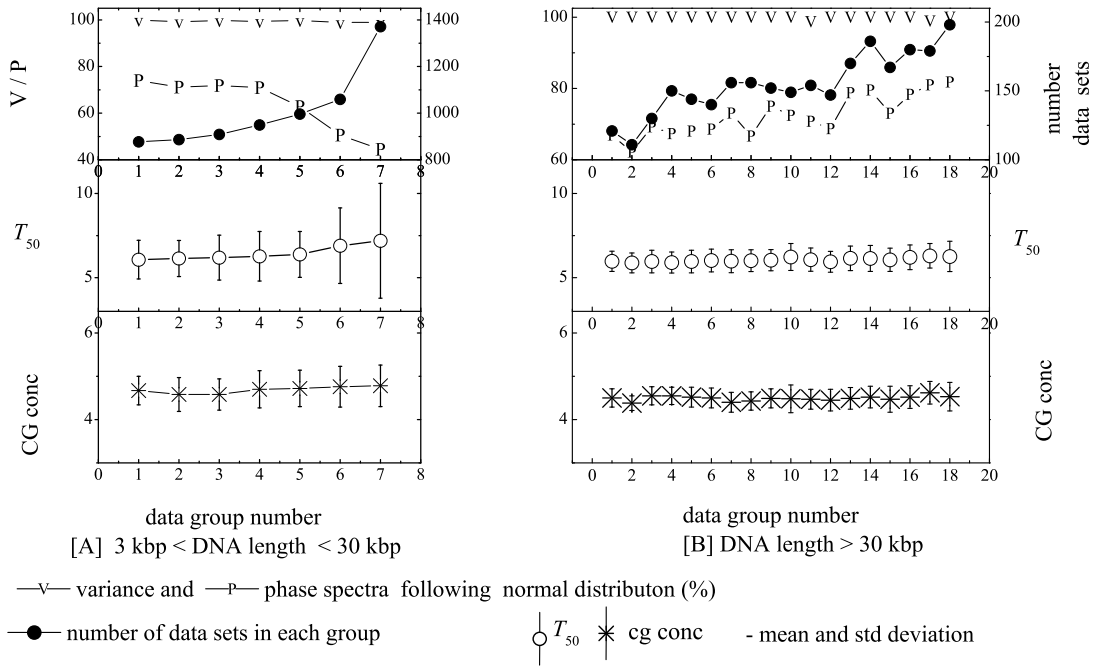

—V— variance and —P— phase spectra following normal distributon (%)

—●— number of data sets in each group     ○ $T_{50}$   ✳ cg conc   - mean and std deviation

Fig. 4. Details of data sets and averaged results of spectral analyses

The variance spectra follow the model predicted and the statistical normal distribution in almost all the data groups in both the data categories, the model predicted spectrum being close to statistical normal distribution for normalized deviation $t$ values less than 2 on either side of $t = 0$ as shown in Sec.3 and Fig. 2 . Li and Holste [77] have identified universal $1/f$ noise, crossovers of scaling exponents, and chromosome-specific patterns of guanine-cytosine content in DNA sequences of the human genome. A total average of about 63.7% and 72% of the data sets (Fig. 4), respectively, in categories A and B show that phase spectra follow the statistical normal distribution.

The average variance and phase spectra along with standard deviations for the data groups in the two categories A and B and the statistical normal distribution are shown in Fig. 5. The average variance spectra follow closely the statistical normal distribution for categories A and B data groups. The average phase spectra for category B data groups alone follow closely the statistical normal distribution while average phase spectra for category A data groups show appreciable departure from statistical normal distribution in agreement with the percentage number of variance spectra (V) and phase spectra (P) following the universal inverse power law form of the statistical normal distribution shown in Fig. 4.

Poland [78,79] reported that C--G distribution in genomes is very broad, varying as a power law of the size of the block of genome considered and they also find the power law form for the C-G distribution for all of the species treated and hence this behavior seems to be ubiquitous.

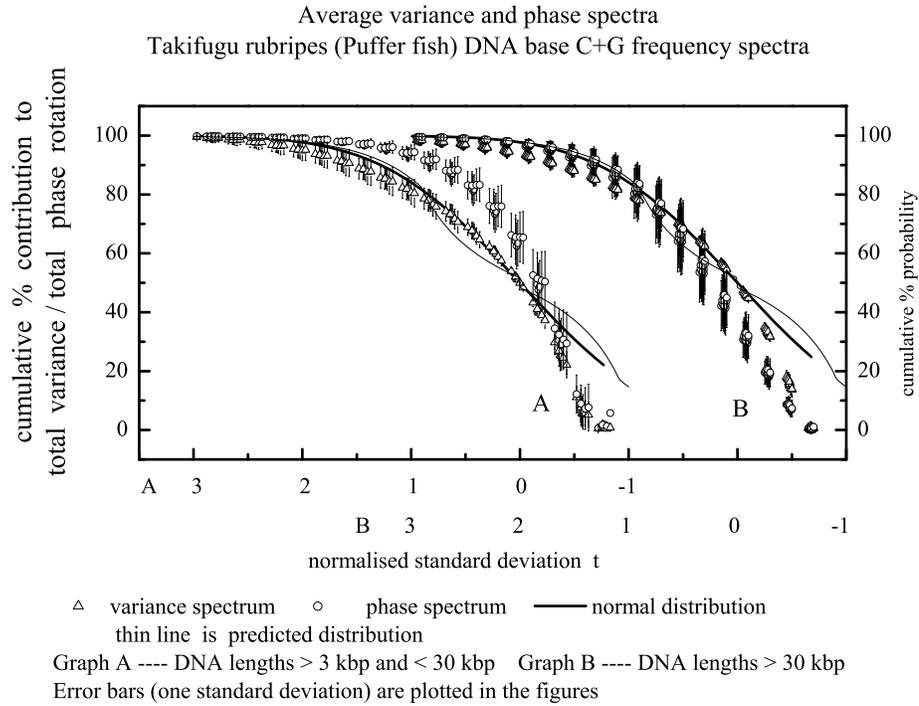

Fig. 5. The average variance and phase spectra of frequency distribution of bases C+G in Takifugu rubripes (Puffer fish) for the data sets given in Figs. 3 and 4. The power spectra were computed as cumulative percentage contribution to total variance versus the normalized standard deviation $t$ equal to $(\log L / \log T_{50}) - 1$ where $L$ is the wavelength in units of 10bp and $T_{50}$ is the wavelength up to which the cumulative percentage contribution to total variance is equal to 50. The corresponding phase spectra were computed as the cumulative percentage contribution to total rotation (Section 2.8).

## 7.2 Model predicted dominant wavebands

Spectral analysis   DNA base CG frequency distribution
Takifugu rubripes (Puffer fish) DNA Release 4
Average distribution of dominant wavebands

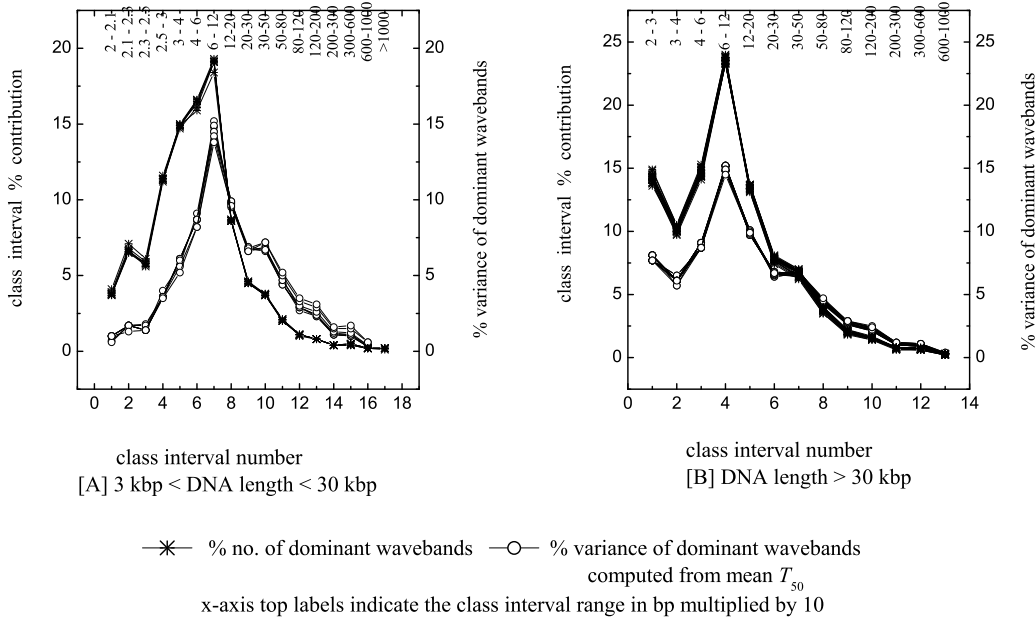

— * — % no. of dominant wavebands   —O— % variance of dominant wavebands
computed from mean $T_{50}$
x-axis top labels indicate the class interval range in bp multiplied by 10

Fig. 6. Dominant wavelengths in DNA bases C+G concentration distribution. Average class interval-wise percentage frequency distribution of dominant (normalized variance greater than 1) wavelengths is given by — * — (line + star). The corresponding computed percentage contribution to the total variance for each class interval is given by — O — (line + open circle). The observed frequency distribution of dominant eddies closely follow the model predicted computed percentage contribution to total variance

The general systems theory predicts that the broadband power spectrum of fractal fluctuations will have embedded dominant wavebands, the bandwidth increasing with wavelength, and the wavelengths are functions of the golden mean (Eq. 2). The first 13 values of the model predicted [3-6] dominant peak wavelengths are 2.2, 3.6, 5.8, 9.5, 15.3, 24.8, 40.1, 64.9, 105.0, 167.0, 275, 445.0 and 720 in units of the block length 10bp (base pairs) in the present study. The dominant peak wavelengths in Category A data sets with shorter DNA lengths (< 30Kbp) were grouped into 17 class intervals 2-2.1, 2.1-2.3, 2.3-2.5, 2.5-3, 3 - 4, 4 - 6, 6 - 12, 12 - 20, 20 - 30, 30 - 50, 50 - 80, 80 – 120, 120 – 200, 200 – 300, 300 – 600, 600 - 1000 (in units of 10bp block lengths) to include the model predicted dominant peak length scales mentioned above. For Category B data sets with longer DNA lengths (> 30Kbp), the dominant peak wavelengths were grouped into 13 class intervals 2 - 3, 3 - 4, 4 - 6, 6 - 12, 12 - 20, 20 - 30, 30 - 50, 50 - 80, 80 – 120, 120 – 200, 200 – 300, 300 – 600, 600 - 1000 (in units of 10bp block lengths) to include the model predicted dominant peak length scales. Category A with shorter DNA lengths will exhibit more number of dominant wavebands in the shorter wavelengths and therefore more number of class intervals in the shorter wavelength region 2 – 3 (in units of 10bp block lengths) for Category A. The class intervals increase in size progressively to accommodate model predicted increase in bandwidth associated with increasing wavelength.

Average class interval-wise percentage frequencies of occurrence of dominant wavelengths (normalized variance greater than 1) are shown in Fig. 6 along with the percentage contribution to total variance in each class interval corresponding to the normalised standard deviation $t$ (Eq. 5) computed from the average $T_{50}$ (Fig. 4). In this context it may be mentioned that statistical normal probability density distribution represents the eddy variance (Eq. 6). The observed frequency distribution of dominant eddies follow closely the computed percentage contribution to total variance.

## 7.3 Peak wavelength versus bandwidth

The model predicts that the apparently irregular fractal fluctuations contribute to the ordered growth of the quasiperiodic Penrose tiling pattern with an overall logarithmic spiral trajectory such that the successive radii lengths follow the Fibonacci mathematical series. Conventional power spectral analysis resolves such a spiral trajectory as an eddy continuum with embedded dominant wavebands, the bandwidth increasing with wavelength. The progressive increase in the radius of the spiral trajectory generates the eddy bandwidth proportional to the increment $d\theta$ in phase angle equal to $r/R$. The relative eddy circulation speed $W/w_*$ is directly proportional to the relative peak wavelength ratio $R/r$ since the eddy circulation speed $W = 2\pi R/T$ where $T$ is the eddy time period. The relationship between the peak wavelength and the bandwidth is obtained from Eq. (1), namely, $W^2 = \dfrac{2}{\pi} \dfrac{r}{R} w_*^2$.

Considering eddy growth with overall logarithmic spiral trajectory

$$\text{relative eddy bandwidth} \propto \frac{r}{R} \qquad (38)$$

The eddy circulation speed is related to eddy radius as

$$W = \frac{2\pi R}{T} \qquad (39)$$

$$W \propto R \propto \text{peak wavelength}$$

The relative peak wavelength is given in terms of eddy circulation speed as

$$\text{relative peak wavelength} \propto \frac{W}{w_*} \qquad (40)$$

Spectral analyses - DNA base CG sequence
Takifugu rubripes (Puffer fish) Release 4

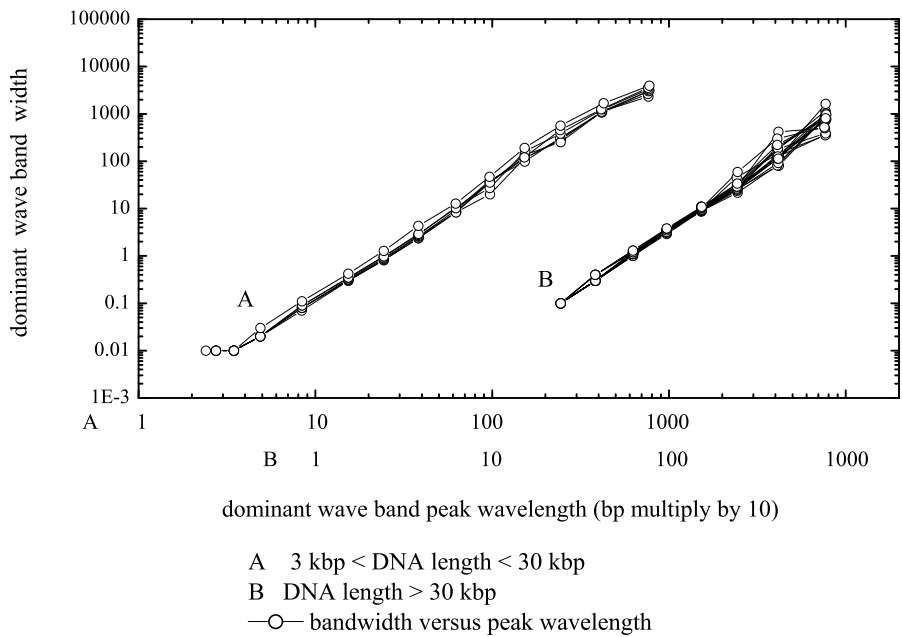

A    3 kbp < DNA length < 30 kbp
B    DNA length > 30 kbp
—o—  bandwidth versus peak wavelength

Fig. 7. Log-log plot of peak wavelength versus bandwidth for dominant wavebands

Spectral analyses - DNA base CG sequence
Takifugu rubripes (Puffer fish) Release 4
Dominant wavebands versus peak wavelength

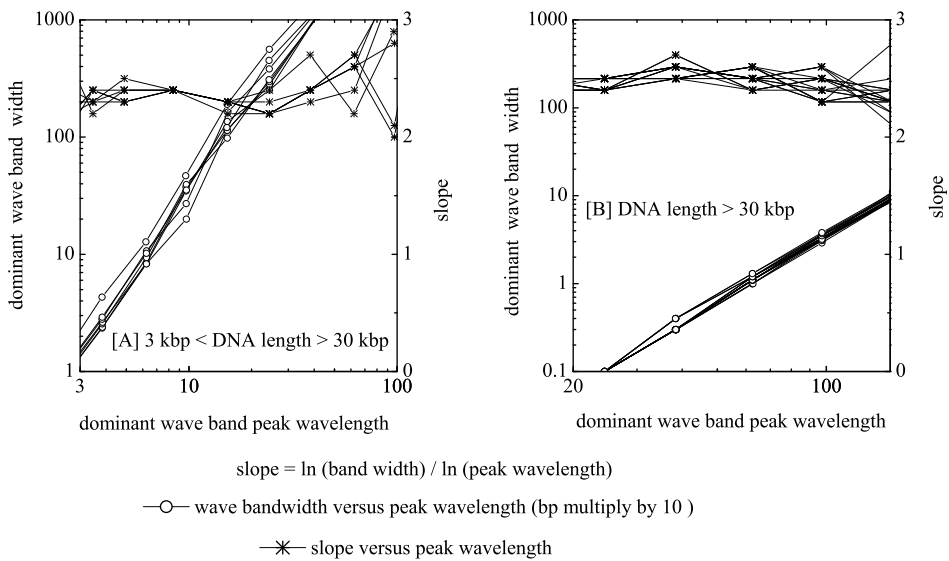

slope = ln (band width) / ln (peak wavelength)

—o—  wave bandwidth versus peak wavelength (bp multiply by 10 )

—*—  slope versus peak wavelength

Fig. 8. Same as in Fig. 6 for a limited peak wavelength range along with the corresponding computed slope
equal to ln (bandwidth) / ln (peak wavelength).

From Eq. (1) the relationship between eddy bandwidth (Eq. 38) and peak wavelength (Eq. 40) is obtained as

$$eddy\ bandwidth = (peak\ wavelength)^2$$

$$\frac{\log(eddy\ bandwidth)}{\log(peak\ wavelength)} = 2 \tag{41}$$

A log-log plot of peak wavelength versus bandwidth will be a straight line with a slope (bandwidth/peak wavelength) equal to 2. A log-log plot of the average values of bandwidth versus peak wavelengths shown in Figs. 7 and 8 exhibit average slopes approximately equal to 2.5.

## 7.4 Quasiperiodic Penrose tiling and packing efficiency

Ten base pairs occur per turn of the DNA helix. Therefore, the fundamental length scale in the case of the DNA molecule is 10 base pairs forming a near complete circle (360 degrees) with 10-fold symmetry. Also, fivefold symmetry is exhibited by the carbon atoms of the sugar-phosphate backbone structure which supports the four bases (A, C, G, T) [63]. The ten-fold and five-fold symmetries underlying the DNA architecture may signify spatial arrangement of the DNA bases in the mathematically precise ordered form of the quasiperiodic Penrose tiling pattern (Fig. 1).

Model predicted universal inverse power law followed by power spectra of CG fluctuations imply spontaneous organization of the DNA base sequence in the form of the quasicrystalline structure of the quasiperiodic Penrose tiling pattern (Fig. 1), characterized by a nested loop (coiled coil) structure. The Fibonacci sequence underlying the quasiperiodic Penrose tiling pattern is found in nature in plant phyllotaxis [18,21]. The model predicted quasicrystalline structure for the linear DNA string is associated with maximum packing efficiency [19] and may help identify the geometry of the compact DNA structure inside the cell nucleus.

The dynamical architecture of the cell nucleus can be regarded as one of the "grand challenges" of modern molecular and structural biology. The genomic DNA and the histone proteins compacting it into chromatin comprise most of the contents of the nucleus. In every human cell, for instance, $6 \times 10^9$ base pairs of DNA-that is, a total length of about 2 meters-must be packed into a more or less spheroid nuclear volume about 10-20 μm in diameter [74]. The reported packing efficiency of the DNA string is then equal to about $10^5$ and is shown in the following to result from 10 stages of successive coiling.

The packing efficiency of the quasicrystalline structure is computed as follows. A length $L$ of DNA sequence in the approximately circular coiled form has a diameter $d$ equal to $L/\pi$ and therefore has a packing efficiency $P_{eff}$ equal to d/L = π. For N stages of successive coiling, $P_{eff}$ is equal to $\pi^N$ and for N=10, $P_{eff} = 10^{4.97} \approx 10^5$, i.e., ten stages of successive coiling results in five-fold compaction of the long DNA string inside the cell nucleus in agreement with reported values [74].

## 8. Conclusions

The power spectra of Takifugu rubripes (Puffer fish) DNA base CG density per 10bp frequency distributions for different DNA lengths follow the model predicted universal inverse power law form which is close to statistical normal distribution for normalized deviation $t$ values less than 2 on either side of mean, signifying model predicted quantumlike chaos or long-range correlations in the spatial distribution of CG concentration in the DNA molecule. Such nonlocal connections enable information communication and control along the total DNA length incorporating coding and noncoding sequences for maintaining robust and optimum performance of vital functions of the living system in a noisy environment. Noncoding DNA sequences are therefore essential for maintenance of functions vital for life. Recent studies show that the proportion of noncoding DNA in the genomic DNA increases with increasing complexity of the organism. The amount of noncoding DNA per genome is a more valid measure of the complexity of an organism than the number of protein-coding genes, and may be related to the emergence of a more sophisticated genomic or regulatory architecture, rather than simply a more sophisticated proteome [65]. One of the steps in turning genetic information into proteins leaves genetic fingerprints, even on regions of the DNA that are not involved in coding for the final protein. They estimate that such fingerprints affect at least a third of the genome, suggesting that while most DNA does not code for proteins, much of it is nonetheless biologically important – important enough, that is, to persist during evolution [80]. Much non-coding DNA has a regulatory role (81, 82).

## Acknowledgement

The author is grateful to Dr. A. S. R. Murty for encouragement during the course of the study.